# ATLAS and CMS applications on the WorldGrid testbed

V. Ciaschini, A. Fanfani, F. Fanzago, M. Verlato, L. Vaccarossa
*INFN, Sez. di Bologna-Padova- Milano, Italy*

F. Donno
*CERN, Geneva, Switzerland*

V. Garbellotto
*Department of Information Engineering, University of Padova, Italy*

WorldGrid is an intercontinental testbed spanning Europe and the US integrating architecturally different Grid implementations based on the Globus toolkit. It has been developed in the context of the DataTAG and iVDGL projects, and successfully demonstrated during the WorldGrid demos at IST2002 (Copenhagen) and SC2002 (Baltimore). Two HEP experiments, ATLAS and CMS, successful exploited the WorldGrid testbed for executing jobs simulating the response of their detectors to physics events produced by real collisions expected at the LHC accelerator starting from 2007. This data intensive activity has been run since many years on local dedicated computing farms consisting of hundreds of nodes and Terabytes of disk and tape storage. Within the WorldGrid testbed, for the first time HEP simulation jobs were submitted and run indifferently on US and European resources, despite of their underlying different Grid implementations, and produced data which could be retrieved and further analysed on the submitting machine, or simply stored on the remote resources and registered on a Replica Catalogue which made them available to the Grid for further processing. In this contribution we describe the job submission from Europe for both ATLAS and CMS applications, performed through the GENIUS portal operating on top of an EDG User Interface submitting to an EDG Resource Broker, pointing out the chosen interoperability solutions which made US and European resources equivalent from the applications point of view, the data management in the WorldGrid environment, and the CMS specific production tools which were interfaced to the GENIUS portal.

## 1. INTEROPERABILITY SOLUTIONS WITHIN THE WORLDGRID TESTBED

The WorldGrid testbed infrastructure is based on the middleware provided by the EDG 1.2 release [1] for European sites and the VDT 1.1.3 release [2] for US sites. It consists of one central EDG-Resource Broker (RB) at INFN-Pisa with a backup at INFN-Milan; one central EDG-Information Index (II) at INFN-Pisa with a backup at INFN-Milan; one central EDG-RB GLUE-compliant at INFN-CNAF (Bologna); one central EDG-Replica Catalogue (RC) at INFN-CNAF; two EDG-User Interfaces (UIs) running the GENIUS portal at INFN-Padova and INFN-Pisa and two UIs running the Grappa portal at ANL and INFN-Milano; one Virtual Organization (VO) LDAP server at INFN-CNAF for DataTAG authorized users, and one at ANL for iVDGL authorized users; 8 European sites hosting EDG Computing Elements (CEs), Worker Nodes (WNs) and Storage Elements (SEs): Bologna, Milano, Padova (INFN-Italy), Valencia (Spain), Geneva (Switzerland), Bristol (UK), Karlsruhe (Germany) and Lisbon (Portugal); 9 US sites hosting VDT servers and clients at Gainesville (FL), Batavia (IL), Bloomington (IN), Boston (MA), Milwaukee (WI), San Diego (CA), Pasadena (CA), Argonne (IL) and Brookhaven (NY). As Local Resource Manager System (LRMS), PBS was adopted by 7 European sites and 5 US sites, LSF by 1 European site and Condor by 4 US sites. The operating systems supported are RH6.2, Fermi 7.1 and RH7.2.

In order to achieve interoperability between the two testbeds, i.e. allow for job submission through the EDG-RB to both European and US sites, additional software has been identified and installed where needed on top of the EDG and VDT distributions.

It concerns:

- VO management tools for authentication/authorization: they are missing in VDT-1.1.3 while already part of EDG-1.2; they allow for the testbed sites to accept certificates issued by EDG and DOE Certification Authorities, and automate the creation of local grid-mapfiles starting from the lists of authorized users published in the VO LDAP servers;
- Information Services: the Globus Schema and Information Providers (IPs), part of VDT-1.1.3, are not descriptive enough for the resource discovering and brokering process performed by the EDG-RB. It is therefore necessary to install and configure the EDG Schema and IPs on the VDT servers together with the existent Globus descriptions. Since EDG-1.2 does not provide IPs for Condor LRMS, the 4 US sites running it could not be selected by the EDG-RB for job execution. By the time of the SC2002 demo, a few of EDG and VDT sites were equipped also with the new GLUE Schema and IPs [3]. A new developed EDG-RB GLUE-compliant has been deployed at INFN-CNAF and its use demonstrated at SC2002;
- Applications: ATLAS and CMS jobs assume to find part of the experiment specific software installed on the WN where they are executed. The experiment software distribution is already included in the standard EDG-1.2, and had to be installed on VDT servers;

- Data Management: some EDG-1.2 tools used by ATLAS and CMS applications for running their jobs and managing their input/output data on EDG testbed are missing in VDT-1.1.3;
- Monitoring: two different monitoring tools, Nagios [4] and VO Ganglia [5] have been adopted, and the corresponding packages installed and configured on all sites.

Since EDG and VDT sites use different packaging methods and tools, namely RPM and LCFG [6] for EDG and Pacman [7] for VDT, the "interoperability" software has been packaged into RPM as well as Pacman caches in order to allow easy installation on every node.

By the time of the IST2002 demo, a total of 8 European sites and 5 US sites were properly set up and could be selected by the EDG-RB for job submission, fully exploiting its resource discovery and match-making capabilities. VDT servers were configured to act as CE and SE at the same time, while VDT clients acted as WN in the EDG language. The WorldGrid testbed can be seen as an homogeneous set of resources where ATLAS and CMS application jobs could run indifferently on European or US side. Moreover, INFN-Milano, INFN-Padova and Gainesville (FL) sites install GLUE IPs too, and can be selected by the GLUE-compliant EDG-RB deployed in time for SC2002 demo.

## 2. JOB SUBMISSION FROM EUROPE THROUGH THE GENIUS PORTAL

The GENIUS portal (**G**rid **E**nabled web e**N**vironment for site **I**ndependent **U**ser job **S**ubmission) [8] is a product developed by the INFN-Italy and NICE SRL, within the INFN-GRID project. It is installed on the EDG-UI and allows users belonging to any VO for: secure authentication and authorized access to the Grid; job submission and job status monitoring on the Grid; monitoring of Grid resources; data management over the Grid (upload, download, replication and publication of files); Replica Catalogue browsing; interactive sessions on remote EDG-UI.

At the beginning of the session, the user is asked to upload his personal Grid certificate to the remote EDG-UI hosting the portal, and create the time-limited proxy allowing for access to the Grid. An user job is simply described by a JDL file (the Job Description Language [9], part of the EDG Workload Management System), which completely defines a job in terms of executable, arguments, input/output, special requirements and so on. The JDL file can be uploaded to or directly edited in the remote EDG-UI. If needed, additional input files can be uploaded to the EDG-UI. Then the JDL file, containing the requirements for running on a Grid resource where ATLAS or CMS software was installed, is submitted to a given EDG-RB. Every EDG-RB is connected to a set of Grid resources through the EDG-II, according to the EDG architecture, and GENIUS allows the user for selecting it among a list of available EDG-RBs. The EDG-RB service, after receiving the job, performs a match-making between the job requirements and the resources characteristics, scheduling the application job only on farms publishing in the Grid Information System their installation of the ATLAS or CMS software and where the Grid user is authorized to run. In the WorldGrid testbed, all demo users were authorized to run all over the participating sites, and all sites have an installation of both certified ATLAS and CMS software, so the job could run everywhere on US or European resources. At the end of the job, output files can be retrieved on the submitting EDG-UI and immediately analysed through the portal via its VNC [10] interface allowing the user to open an interactive graphic session on the EDG-UI.

Every VO can adopt this approach, once the VO-specific job description can be translated into a JDL file. Nevertheless, the GENIUS portal allows for a more VO-specific customisation, defining the so-called "VO Services" item. This is the case of CMS, which has since time a set of Monte Carlo production tools (IMPALA, BOSS, RefDB) developed for automating the simulation job submission and monitoring on local farms. These tools were at first adapted in order to be used in the EDG environment, and then in a subsequent step interfaced to the GENIUS portal. GENIUS allows for a user-friendly configuration and use of these tools from any remote web browser, once they are installed on the EDG-UI hosting the portal. The EDG-UIs in INFN-Padova and INFN-Pisa used in the WorldGrid testbed, were equipped with these additional CMS production components and the GENIUS portal.

## 3. ATLAS APPLICATION JOB SUBMISSION

The ATLAS Collaboration [11] at CERN is preparing for the future data taking and analysis, by running the so-called Data Challenges (DCs) [12]: a set of computing tasks, involving software simulation of physics processes to be studied. The goals of the ATLAS DCs are the validation of the Computing Model, the complete software suite and the data model, and furthermore to ensure the correctness of the technical choices to be made. It is understood that these DCs should be of increasing complexity and will in future use the software which will be developed in the LHC Computing Grid (LCG) project [13], to which ATLAS is committed, as well as the Grid middleware being developed in the context of several Grid projects and toolkits like EDG and VDT. The results of these data challenges will be used as an input for the Computing Technical Design Report and in preparation of the Computing Memorandum of Understanding in due time.

Given the volume of the data to be produced and analysed, ATLAS distributes the whole Data Challenges task among the collaborators over the world (150 universities and laboratories in 34 countries). In the DC1 stage, 15 countries from USA to Norway to Japan to Australia participated. It is natural that with such a spread of resources, ATLAS is eager to make use of the Grid tools in order to enable a distributed production infrastructure. Some part of the first phase of the DC1 was performed using tools developed by various Grid projects (e.g., the NorduGrid [14] in Denmark, Norway and Sweden, VDT in USA). It was planned to perform the second phase of the DC1 using Grid tools as much as reasonably possible.

This contribution describes the extension of the ATLAS production environment to the WorldGrid testbed and its interface to the GENIUS portal.

### 3.1. ATLAS DC1 job description

DC1 is a use case perfectly suited to be executed in the Grid environment. The job flow is fairly simple, and is representative of a typical data processing not only in High Energy Physics, but also in many other research areas.

The task of the ATLAS DC1, phase 1, was to process several millions of previously generated events by applying the detector simulation software. The output consists of new data, describing each processed event in terms of the ATLAS detector electronics response. Hence, the processing chain is as follows: input file ? detector simulation ?  output files. Each input file results in several output ones.

To validate the WorldGrid testbed a short (first 100 input events) DC1 job was submitted to each site advertising ATLAS-3.2.1. The output of this job is known, and the fact that all the sites produced identical results was considered to be a sufficient proof of the proper ATLAS software installation.

Input for the Data Challenge 1 is a set of PYTHIA [15] generated events, stored as ROOT [16] files, known as "Input Partitions". Input data are processed by the ATLAS detector simulation software. It was chosen to pre-install all the necessary software prior to task execution at all involved sites.

A full event simulation takes about 150 seconds per event on a typical processor of 1 GHz. Every job produces a set of output files: simulated events are stored in ZEBRA files ("Output Partitions"), the control distributions are saved in HBOOK files. Standard output and standard error files constitute the job log.

### 3.2. ATLAS Event Simulation

The ATLAS detector simulation was done in the ATLSIM framework using GEANT3 [17]. ATLSIM is a PAW-based framework, which uses KUIP instead of FFREAD for job control. It has an improved memory management, eliminating any hard limits on the track/vertex/hit numbers. It also has improved hadronic physics based mainly on the GCALOR package. A number of known infinite loops were eliminated. To avoid known problems with low energy $K_{0L}$ (zero cross section in FLUKA), they are always traced by GEISHA.

ATLSIM uses plug-in components (shared libraries) to provide extra I/O facility (ROOT, Objectivity) and to load ATLAS detector geometry. Description of the ATLAS geometry is taken from the DICE package.

During the simulation phase di-jet events produced by PYTHIA where analysed by a filtering routine which looked for a predefined energy deposition in two neighbouring towers in $\eta-\phi$ space. Only events selected by the filter were passed to the simulation step and then written out.

### 3.3. ATLAS Software distribution

The ATLAS software source code is maintained at CERN in a CVS repository and then installed and compiled in a public AFS directory, under the ATLAS tree. The compilation process is done on Linux machines running CERN RedHat Linux 6.1. Users may use executables and the data files directly linking them from CERN. This approach is anyway not indicated for remote sites with a bad connection to CERN or without access to AFS. For this purpose a set of RPM packages has been produced, in order to install the full ATLAS software distribution on machines both with and without AFS. The RPM kit [18] has been designed to be used on standard as well as on EDG machines, in order to fulfill the requirements of the DC1.

The ATLAS software release 3.2.1 RPM suite has been proven to be robust and efficient during the first phase of the ATLAS DC1, in which several millions of events have been successfully simulated on different machines and Linux distributions, these RPMs were distributed and installed as a part of the EDG toolkit, to provide the ATLAS runtime environment ATLAS-3.2.1.

Starting from the RPM suite, a distribution suited to be installed by Pacman was made and put on the ScienceGrid cache [19] to allow installation over US VDT resources. It included also additional components (lesstif, zsh) needed at run-time, which were missing on VDT servers, and an installation check package.

### 3.4. Job workflow and data management

An ATLAS user, after authenticating to the Grid via the GENIUS tool, uses the GENIUS File Service to upload on the remote EDG-UI hosting the portal the JDL file prepared for the DC1 ATLAS job, together with some needed input files that will be shipped to the WN via the InputSandbox mechanism. Then he/she selects the GENIUS Job Submission Service option and picks up the JDL file to be submitted to the previously selected WorldGrid Resource Broker. The RB queries the Information System obtaining the list of resources available to satisfy the job requirements (mainly the ones having the ATLAS software installed), and the user can select the resource where the job will be run, or let the RB choose the best matching resource according to optimisation algorithms specified in the JDL file through the JDL Rank attribute (the default Rank foresees the selection of the Computing Element with the minimum estimated traversal time). Fig.1 shows the screen snapshot of the job submission step. When, after the submission, the job reaches the CE, it enters the queue of the LRMS running on the CE (PBS or LSF are required), and then it is dispatched to a WN. The InputSandbox files are copied using the GridFTP protocol from the RB to the WN, and then the ATLAS job starts its execution. When the job ends, output data and log files are copied back to the RB, and will be retrieved asynchronously in the EDG-UI user home directory upon the user's request. The Workload Management System manages Input and Output Sandboxes transfers. The output data files include a control histogram file used to validate the job result. After the output retrieval, this file can be graphically analysed remotely via a PAW session in the EDG-UI exploiting the GENIUS Interactive Service, as shown in Fig.2.

## 4. CMS APPLICATION JOB SUBMISSION

The CMS Monte Carlo production environment already allows, on CMS-dedicated local farms (Regional Centres), for a large-scale production process minimizing at the same time the human intervention [20]. It consists of automated subsystems taking care of input parameter management, robust and distributed request and production accounting, job preparation and submission to the batch system, job tracking and bookkeeping, data and replica management. The original environment was modified in order to adapt it to the provided EDG middleware tools. The first tests of integration of CMS production tools with EDG were performed on a dedicated DataTAG-CMS testbed, where on top of the EDG middleware CMS specific packages were installed on the CEs/WNs. They became part of the EDG-1.2 release and were installed automatically on the whole European EDG testbed.

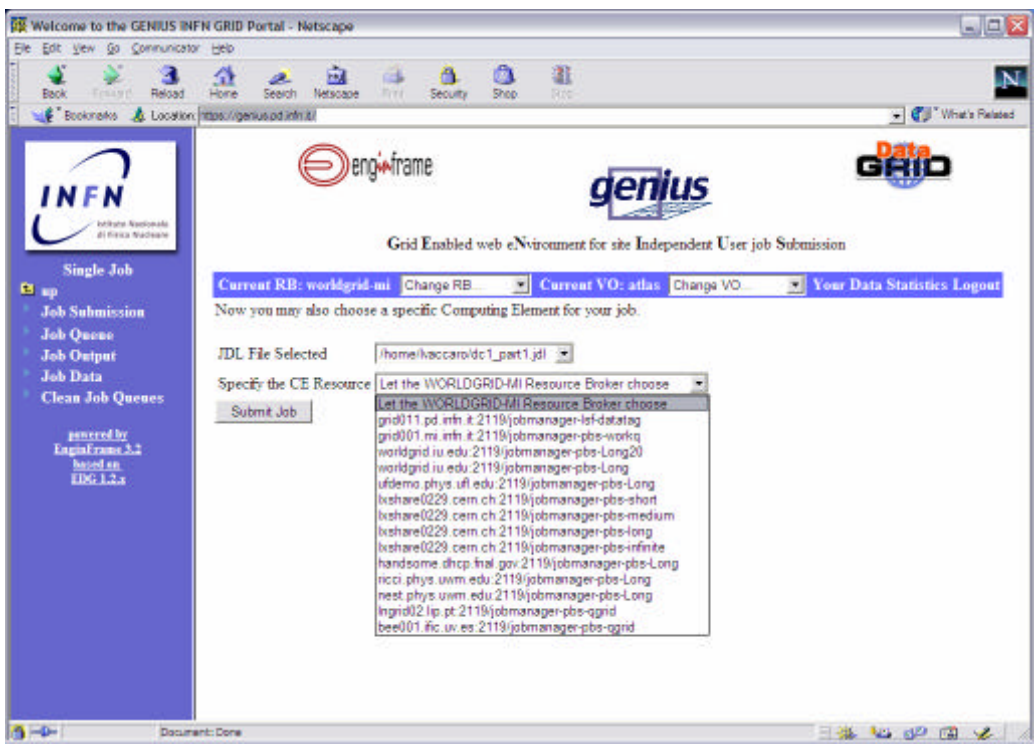

Figure 1: Submission of an ATLAS job through the GENIUS Job Submission Service

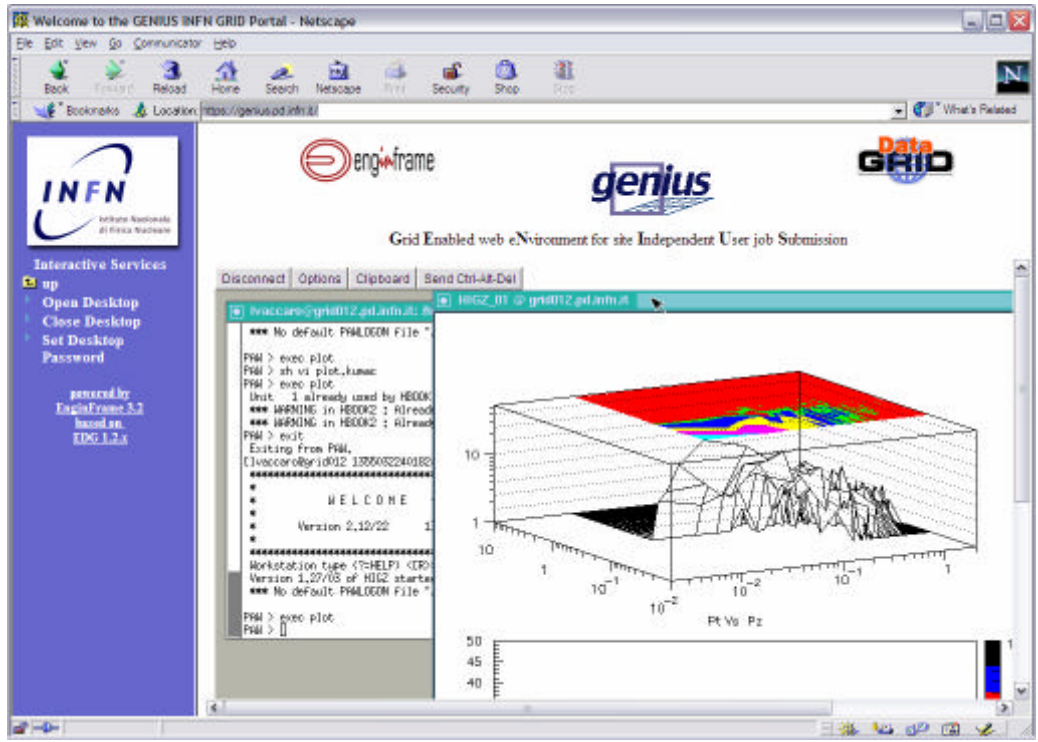

Figure 2: Remote graphical analysis of the ATLAS control histograms through the GENIUS Interactive Service

This contribution describes the extension of the CMS production environment to the WorldGrid testbed and its interface to the GENIUS portal.

### 4.1 CMS Monte Carlo production chain

The CMS data simulation chain consists actually of 4 steps:
- CMKIN: generation of physics events producing ntuples (random access Zebra files) with the 4 vectors of the particles generated in the physics collisions. CMKIN is a statically linked FORTRAN program based on the PYTHIA generator. CMKIN requires some control cards as input to define the physical process to be simulated and a CPU time of the order of half a second on a 1 GHz PIII CPU to generate a single event. The output file size is about 50 KB/event.
- CMSIM: tracking simulator in different components of the CMS detector, producing output FZ files (sequential access Zebra files). CMSIM is a statically linked FORTRAN program based on GEANT3 that takes as input the ntuple produced in the CMKIN step and some additional control cards. The CPU time to generate a single event is of the order of 350 seconds on a 1 GHz PIII CPU, and the output file size is about 1.8 MB/event.
- OOHIT formatting: collection and conversion of CMSIM output files into an Objectivity/DB database through ORCA [21], a dynamically linked C++ program using Objectivity/DB as underlying persistency layer;
- Digitisation: superimposition of pile-up events to the signal sample and final simulation of the response of the CMS data acquisition system using ORCA program.

The first two steps are suited to be run widespread over the Grid, while the last two require licensed software and could be run on dedicated local farms only. The integration effort with EDG middleware and the extension to the WorldGrid testbed concerned the CMKIN and CMSIM steps only. Nevertheless the CMSIM output data produced on the WorldGrid, even if limited in quantity at small test samples, were at the same quality level of the official CMS productions. CMKIN and CMSIM programs were included in a set of RPMs composing the EDG-1.2 release. Therefore they were already installed on the European resources of the WorldGrid, while a CMS Pacman cache was set up in order distribute the software over the US VDT resources.

## 4.2 CMS Monte Carlo production tools

Each step of the production chain described above is managed in the CMS production environment by the following tools:

- RefDB [22] is a SQL database located at CERN storing the production requests. Each request contains all parameters needed for producing a given physics data sample (dataset). RefDB allows for the registration of production requests through a web interface and the retrieval of the corresponding "assignment ID" together with all relevant parameters to be given as input to the IMPALA component. RefDB also stores the summary information about the assigned production sent back by a Regional Centre. It guarantees in this way a high-level overall coordination. No modifications were needed to allow for RefDB use on the Grid environment, since all interactions with the whole system happens through its interface with the IMPALA tool;
- IMPALA [23] is a set of scripts that, taking as input a RefDB assignment, retrieve from the central database all needed parameters (Job Declaration step), generate accordingly instances of jobs from templates (Job Creation step), submit the jobs to a job scheduler (Job Submission step). It also provides a system to trace the status and the progress of the job. This tool was heavily modified in order to integrate it with EDG, for example preparing JDL file in addition to the job script and adapting the job to run in a EDG environment. The EDG-enabled version of IMPALA was installed on the EDG-UIs. The same modifications were of course valid also when extending to the WorldGrid testbed, where the main effort instead was focused to interface IMPALA with the GENIUS portal. In fact, even the modified IMPALA version had a simple command line interface for issuing job declaration, creation and submission commands, after having properly customized a few configuration files. A GENIUS CMS-VO Service called IMPALA was set up using scripts (bash and PERL) and XML files, allowing users to select assignment IDs, define the configuration settings, and execute the IMPALA commands through a friendly web interface with edit/combo boxes and buttons, as shown in Fig.3.
- BOSS [24] provides real-time monitoring and bookkeeping of jobs submitted to a computing resource, storing the information persistently in a relational database (MySQL in the current version) for further processing. It wraps production jobs in order to extract specific job information to be monitored parsing the standard input, output and error of the running job, updating at run-time the database. This step requires WNs with outbound connectivity, as was the case for the WorldGrid testbed. For a given CMS specific job type, like CMKIN or CMSIM, the description of the parameters to be monitored and the scripts for extracting their values are part of the IMPALA software. BOSS is also a submission system, with interfaces to many LRMSs like PBS, LSF and Condor. In order to allow for job submission to the EDG testbed, a new interface was developed to the EDG-RB, which is treated as a new scheduler. Actually, the BOSS software is strictly connected with the IMPALA software, and the IMPALA Job Submission step occurs via this BOSS interface. In the WorldGrid testbed BOSS was installed together with IMPALA on the EDG-UIs. BOSS was interfaced to the GENIUS portal developing a CMS-VO Service called "Status from BOSS", which allows for querying the BOSS database for job monitoring, as shown in Fig.4.

Figure 3: IMPALA interface to GENIUS customizing job preparation and submission

Figure 4: BOSS interface to GENIUS for monitoring every production step

**TUCP004**

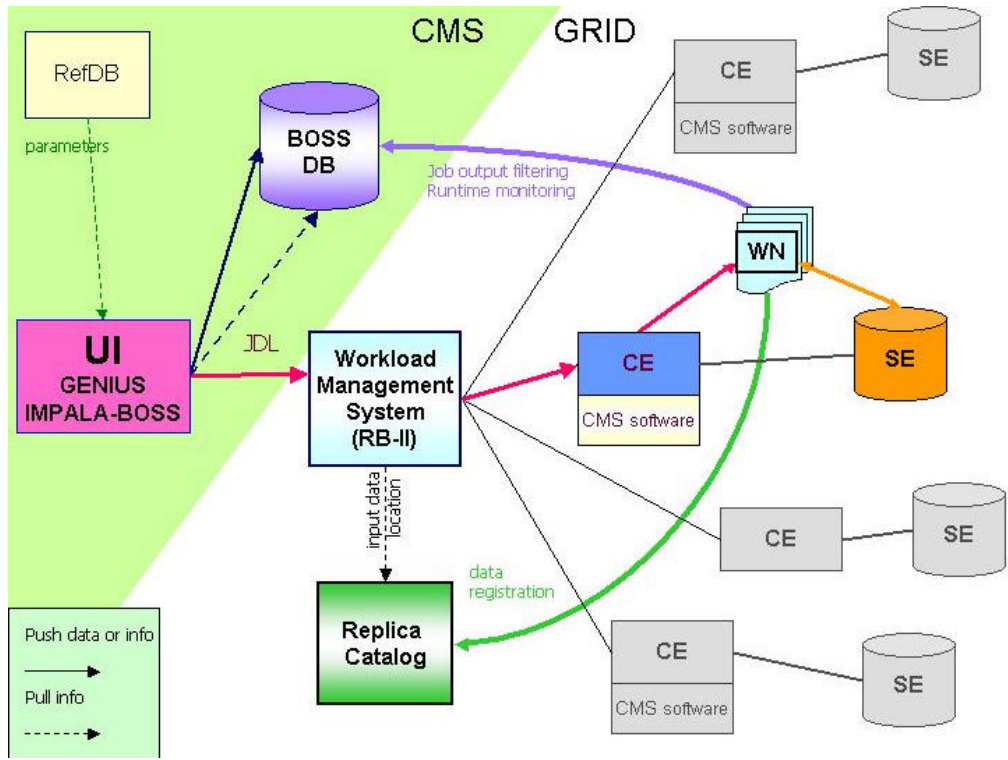

Figure 5: Job workflow schema describing the integration of the CMS production software and tools with the Grid architecture

### 4.3 Job workflow and data management

This section describes the job submission from the GENIUS portal, the job workflow inside the Grid and the data management solutions illustrated in fig.5.

First of all a CMS user has to authenticate to the Grid via the GENIUS tool and to obtain a RefDB assignment ID for the dataset he/she wants to produce. The IMPALA Service in GENIUS is then used to choose the production step (CMKIN or CMSIM) and to perform the IMPALA steps for job creation and submission (Declare, Create and Run the job). At each step all needed information to configure IMPALA - such as the RefDB assignment ID, the number of jobs, the number of events per job, the Resource Broker to be used, the default storage area, etc… - are inserted through the web interface of the IMPALA Service. The JDL files created by IMPALA for the given dataset are automatically processed by BOSS and submitted to the chosen Resource Broker. Some control cards and a few KBytes of input files are shipped to the WN via the InputSandbox mechanism, while the job log files are retrieved back on the UI via the OutputSandbox. The job workflow after the submission to the Workload Management System depends on the production step.

In the CMKIN case, the RB queries the Information System looking for available CEs matching the job requirements (mainly the CMS software must be already installed on the CE publishing this information), and choosing the one with the minimum estimated traversal time. The job is submitted to the CE, i.e. it enters the queue of the LRMS running on the CE (PBS or LSF are required), and then dispatched to a WN. The InputSandbox files are copied using the GridFTP protocol from the RB to the WN, and then the CMKIN user job starts. During job execution some CMS specific monitoring information are sent to the BOSS database server, running on the UI, allowing for run-time production monitoring. When the job ends, CMKIN log files are copied back to the RB and they will be retrieved asynchronously upon UI request. The output data ntuples are copied to the SE "close" (in the LAN sense) to the CE and registered to the Replica Catalogue (i.e. a logical file name is associated to the physical location of the file) to allow for further processing by the CMSIM step. The Workload Management System manages Input and Output Sandbox transfers, while data transfers between WN and SE and registration to the Replica Catalogue are performed inside the job.

In the CMSIM case the scheduling is data driven, the RB queries both the Information System and the Replica Catalogue, looking for available CEs and the physical location of the input CMKIN ntuples to be processed. The job is sent by the RB to a CE close to the SE where the physical ntuples were previously stored. Before the CMSIM execution on the WN, the ntuples are copied from the SE to the WN, as well as the InputSandbox files as in the CMKIN case. During job execution, the BOSS database is updated with CMSIM specific information, and at the end of the job the output data files are copied to the close SE and registered to the Replica Catalogue, while the log files are transferred through the OutputSandbox for asynchronous access as in the CMKIN case.

Anytime during the production, the BOSS database can be queried through the GENIUS Interface, allowing for run-time monitoring of the production process. The CMSIM output data stored in the Grid will be then copied to some sites, outside the Grid, equipped with the proper software and licenses to run further steps of the production chain described in section 4.1.

## 5. CONCLUSIONS AND PERSPECTIVES

To the best of our knowledge, WorldGrid is the first large-scale testbed that combines middleware components and makes them work together. It represents an important step towards interoperability of Grid middleware developed and deployed in Europe and in the US. Hundreds of ATLAS and CMS simulation jobs were submitted and executed successfully, exploiting both European and US computing resources, despite of their underlying different Grid architecture. In addition, a significant amount of jobs were submitted exploiting the newly commissioned GLUE-compliant EDG-RB. The LHC Computing Grid (LCG) first Pilot service deployed in February 2003 [25] has adopted some solutions developed in WorldGrid.

## Acknowledgments

The present work would not have been possible without the friendly collaboration and the joint effort of: S.Andreozzi, S.Fantinel, A.Ghiselli, M.Mazzucato, D.Rebatto, G.Tortone, C.Vistoli (DataTAG project); R.Gardner, J.Gieraltowski, R.Pordes, J.Rodriguez, S.Youssef (iVDGL project); R.Barbera and A.Falzone (GENIUS team); all involved site-administrators.

This research was partially funded by the IST Programme of the European Union under grant IST-2001-32459 (DataTAG project).